%% file: main.tex
\RequirePackage{fix-cm}
\documentclass[smallextended]{svjour3}       
\smartqed  
\usepackage{graphicx}

\usepackage[hidelinks]{hyperref}

\usepackage{xcolor}
\usepackage{amsmath} 
\usepackage{pdflscape}
\usepackage{amsmath}
\usepackage{amsfonts}
\usepackage{natbib}
\usepackage{bm} 
\usepackage{url}

\newcommand{\yd}[1]{\textcolor{cyan}{#1}}
\newcommand{\review}[1]{\textcolor{black}{#1}}

  \renewcommand{\emph}[1]{\textit{#1}}

\begin{document}

\title{\review{Computation and stability analysis of periodic orbits using finite differences, Fourier or Chebyshev spectral expansions 
in time}}


\titlerunning{Stability analysis of periodic orbits using finite differences or spectral expansions}        

\author{Artur Gesla \and Yohann Duguet \and Patrick Le Qu\'er\'e \and Laurent Martin Witkowski}


\institute{
A. Gesla \at
Sorbonne Universit\'e, F-75005 Paris, France\\
\review{LISN-CNRS, Universit\'e Paris-Saclay, F-91400 Orsay, France}
\and
Y. Duguet \at
LISN-CNRS, Universit\'e Paris-Saclay, F-91400 Orsay, France
\and P. Le Qu\'er\'e \at
LISN-CNRS, Universit\'e Paris-Saclay, F-91400 Orsay, France
\and L. Martin Witkowski \at
Universite Claude Bernard Lyon 1, CNRS, Ecole Centrale de Lyon, INSA Lyon, LMFA, UMR5509, 69622 Villeurbanne, France
}

\date{Received: \today / Accepted: date}

\maketitle

\begin{abstract}

\review{
We analyse and compare several algorithms to compute
numerically periodic solutions of high-dimensional dynamical systems and investigate their Floquet stability without building the monodromy matrix. The solution and its perturbation are discretised in time either using finite differences, Fourier-Galerkin or Chebyshev expansions.  The resulting nonlinear set of equations describing the periodic orbit is solved using a Newton-Raphson algorithm. The linearised equations determining the stability lead to a generalised eigenvalue problem. Unlike the Fourier-Galerkin method, the use of Chebyshev polynomials or finite differences has the advantage that the relevant Floquet exponents are directly given without the well known issue of having to sort out the eigenvalues. The speed of convergence of these three methods is illustrated with examples from the Lorenz system, the Langford system and a two-dimensional thermal convection flow inside a differentially heated cavity. This last example demonstrates the potential of the newly proposed Chebyshev expansion for large-scale problems arising from the discretisation of the incompressible Navier-Stokes equations.}
\end{abstract}

\input{p1}
\begin{acknowledgements}
The authors wish to thank Daniel Henry, Wietze Herreman, Simon Kern, Arnaud Lazarus and Laurette Tuckerman for fruitful discussions and many valuable comments.
\end{acknowledgements}

%
\section*{Conflict of interest}

The authors declare no conflict of interest.

\section*{Data availability statement}

Data sets generated during the current study are available from the corresponding author on reasonable request. 

\bibliographystyle{apalike}      


\bibliography{biblio}


\end{document}

%% file: p1.tex
\setcounter{tocdepth}{4}
\setcounter{secnumdepth}{4}








\section{Introduction}

Identification of periodic orbits (POs) of 
nonlinear dynamical systems, together with the computation of their stability, is crucial in many domains of physics and engineering. 
\review{Periodic solutions are often the simplest 
unsteady solutions and emerge, for instance,
from the local destabilisation of a fixed point solution of the system.}
 In dissipative dynamical systems featuring attractors, periodic orbits can be stable or unstable. Deterministic chaotic attractors are known to be organised around an infinity of unstable periodic orbits ranked by their stability \citep{cvitanovic2005chaos}, which makes periodic orbits and their stability characteristics crucial  
 also in chaotic settings. 

Many numerical approaches to the determination of periodic orbits in ordinary differential equations (ODEs) exist. Popular choices
include shooting methods and their variants (see section \ref{perorb}) and delayed feedback methods 
\citep{pyragas1992continuous}. 
\review{
} 
\review{
Other methods include adjoint-based methods \citep{lan2004variational,farazmand2016adjoint,azimi2022constructing,parker2022variational}.} 
\review{One of the most widely used alternative
methods is a Fourier-based Galerkin method
known as Harmonic Balance Method (HBM)(e.g. \cite{cochelin2009high}).}
HBM 
approximates the periodic orbit as a truncated Fourier series.  
Due to the spectral expression of the cycle, high accuracy or equivalently fast convergence can be achieved. Historically, HBM was used mainly for
low-dimensional ODE systems in mechanical and electrical engineering. Recent studies have considered such methods for higher-dimensional systems stemming from the spatial discretisation of partial differential equations (PDEs),
in particular from fluid mechanics \citep{bengana2019bifurcation,rigas2021nonlinear, SierraAusin_cmame_2022,gesla2023subcritical}.
The linear stability of an HBM solution is usually analyzed by forming the so-called Hill matrix and analyzing its eigenvalues in the search for unstable Floquet exponents. As pointed out by several authors (see e.g. \cite{lazarus2010harmonic}), a
major
drawback of using Hill's matrix is that only a subset of the computed eigenvalues are relevant. As a consequence,  it is necessary to
filter out 
\review{the spurious exponents from the relevant ones.}
 \review{
Various strategies
have been proposed either in}
low-dimensional  \citep{lazarus2010harmonic,peletan2013comparison,wu2022robust,bayer2023sorting} or high-dimensional systems~\citep{pier2017linear,kern2024floquet}. 
\\
\review{The Fourier-Galerkin method referred to here combines the Harmonic Balance method for the search of the periodic solution in combination with Hill's method to investigate its stability.} As the key idea of this article, it is possible to use another 
type of temporal discretisation than the Fourier basis in the HBM equations. \review{The first example described here corresponds to a finite difference discretisation.}
The present work furthermore proposes
a description where the whole periodic orbit and its perturbation are
expanded as a weighted sum of 
Chebyshev polynomials. The stability of the orbit emerges naturally from a generalised eigenvalue problem which involves the Jacobian matrix already computed in the Newton-Raphson method to find the orbit itself, which yields the most unstable Floquet multipliers. 

Several studies have already made use of Chebyshev polynomials in the context of periodic orbits. The earliest use of Chebyshev polynomials for the computation of periodic motion concern orbital motion \citep{kolenkiewicz1967periodic}. \cite{zhou2001research} also used a Chebyshev expansion of a PO in forced and an unforced problems. \cite{zhou2013finding} uses 
the
Chebyshev expansion to approximate the periodic orbit but the stability properties of the orbit are analyzed by forming a monodromy matrix. The authors claim higher accuracy when using a Chebyshev basis compared to a Fourier basis. 
\review{The examples reported hereafter do not support this claim.} A previous study based on a Chebyshev expansion for the linear stability analysis (LSA) of periodic orbits is the work by~\cite{woiwode2023chebyshev}. The authors used Chebyshev polynomials to construct the monodromy matrix built on top of the orbit expressed with Fourier modes,  which differs from the method presented here. 
They also report that the HBM is usually superior to the shooting methods when a smooth periodic solution is considered, yet it suffers from an excessive number of modes in the decomposition in the presence of sharp gradients. It is the aim of the present paper to revisit the use of Chebyshev polynomials in a few examples of dynamical systems with periodic solutions, in comparison with other more standard approaches.




\review{The article is structured as follows. Section~\ref{perorb} 
\review{recapitulates the theoretical basis concerning the periodic solutions of the dynamical systems and the Floquet theory used to study the stability of these solutions.} The different numerical techniques presented include
finite differences, 
 the Harmonic Balance Method,
\review{and a} method 
\review{
based on} Chebyshev expansion.
Section~\ref{examples} presents examples of periodic orbits in the Lorenz and Langford systems, together with the linear stability analysis of these orbits. 
A direct application to 
a periodic solution in a high dimensional system of a fluid flow inside a differentially heated cavity is shown in section \ref{sec: dhc}.
Finally, section~\ref{summary} summarises the presented work.
%
}



\section{Numerical identification of a periodic orbit \review{and stability analysis}} \label{perorb}
\subsection{General concepts }

We consider a
nonlinear system governed by the following equation :
\begin{equation} \label{dyn}
\frac{d\bm{x}}{dt^*}=\bm{f}(\bm{x}), 
\end{equation}
where $\bm x\in\mathbb{R}^d$ is the state vector of the system and {($d\in\mathbb{N^*}$)}
is the number of degrees of freedom, $t^*\in \mathbb{R^+}$ and $\bm f:\mathbb{R}^d\rightarrow \mathbb{R}^d$ is a smooth enough 
function of $\bm x$.
%
A periodic orbit of the system 
satisfies the relation:
\begin{equation} \label{eqperiod}
 \bm{x}(t^*+T)=\bm{x}(t^*)
\end{equation}
for a 
value of $T>0$ and for all times $t^*$. The smallest value of $T$ defines the period of the PO.
\review{Eq. \eqref{eqperiod} does not define the solution in a unique way as any solution shifted in time is also a solution.
 Imposing a phase condition selects one solution (see chapter 10 in \cite{kuznetsov1998elements}).
 } In the current work the time derivative of one of the components of $\bm x $ will be fixed to zero to form such a phase condition. 

\subsection{\review{Basic elements of Floquet theory}}

\review{
Let ${\bm x_p}(t^*)$ be a periodic solution of \eqref{dyn}. We consider a perturbation ${\bm x'}$ to $\bm x_p$ in the form ${\bm x}(t^*)={\bm x_p}(t^*)+{\bm x'}(t^*)$. Linearising \eqref{dyn} around $\bm x_p$ leads to the non-autonomous linear O.D.E. 
\begin{equation} \label{linstab}
\frac{d\bm{x}'}{dt^*}=\bm{A}(t^*)\bm{x}',
\end{equation}
where the linear operator ${\bm A}(t^*)=\frac{\partial {\bm f}}{\partial {\bm x}}(t^*)$ is the Jacobian matrix associated with ${\bm f}$ evaluated along the periodic solution. This operator is time-periodic with the same period as ${\bm x_p}$. 
The stability of the nonlinear solution $\bm x_p$ to infinitesimal perturbations is directly related to the spectral properties of $\bm A(t^*)$. According to Floquet theory 
\citep{floquet1883sur}, in finite dimension $d$,
a general solution ${\bm x'}$ to \eqref{linstab} can be written under the form
\begin{equation} \label{floquet1}
\bm{x}'(t^*)=\sum_{i=1}^{d} c_i {\bm p}_i(t^*)e^{\mu_i t^*},
\end{equation}
where the $\bm p_i$'s are periodic functions of $t^*$  of period $T$,
and the $c_i$'s and $\mu_i$'s are complex numbers. The $d$ exponents $\mu_i,~i=1,...,d$
are by definition the Floquet exponents of ${\bm x_p}$ and they are defined modulo 
$\frac{2\pi}{T}i$. Their real parts, interpreted as Lyapunov exponents, determine the stability of the periodic solution ${\bm x_p}$. In particular, ${\bm x_p}$ is unstable in the Floquet sense if and only if at least one of the Floquet exponents has a strictly positive real part. Alternatively one may consider the characteristic Floquet multipliers $\sigma_i=e^{\mu_iT}$, which are uniquely defined. The orbit is then unstable if the modulus of its largest characteristic multiplier is larger than unity, i.e. $\max{|\sigma_i|}>1$}.\\


\review{In practice, for the computation of the Lyapunov exponents,} 
{consider ${\bm X}(0)={\bm x'_1},..,{\bm x'_d}$ an orthonormal basis at $t^*=0$. For each vector ${\bm x'_i},~i=1,...,d$, integrating 
the linear O.D.E. \eqref{linstab} over exactly the period of ${\bm x_p}$ leads to
the 
monodromy matrix ${\bm X(T)}=[{\bm x'_1}(T),..,{\bm x'_d}(T)]$. Exploiting the periodicity of the ${\bm p}_i's$, we have ${\bm X}(T)={\bm X}(0)e^{RT}$,
with $R$ a time-independent operator. The Lyapunov exponents are directly extracted as the real part of eigenvalues $\mu_i=1,...,d$ of $R$,  ranked by decreasing value \citep{fairgrieve1991ok,chicone2006ordinary}. 
In practice however, for large $d$, due to the cost of accurately integrating $d$ times \eqref{linstab}, forming the monodromy matrix can be computationally prohibitive both in terms of CPU and 
storage. We are hence looking for a way to extract all the leading Floquet exponents without directly forming the monodromy matrix.
}\\


\subsection{\review{General background}}
There is no general way to find all periodic orbits of a nonlinear dynamical system~\citep{kuznetsov1998elements}. 
The simplest method, if a stable PO exists, is to use a numerical time integrator
and solve an initial value problem (IVP). Provided the initial point belongs to the basin of attraction of the requested PO, numerical time integration will converge towards this specific PO.
However for weakly stable orbits convergence may be slow, while unstable POs will not be found using a numerical time integrator alone.\\


If time is rescaled such that the PO is defined over a time interval of length unity, the period $T$ becomes a parameter of the system and finding a PO can be recast as a boundary value problem (BVP).  Equations~\eqref{dyn} and~\eqref{eqperiod} become
\begin{equation} \label{dynrescaled}
\frac{d\bm{x}}{dt}=T \bm{f}(\bm{x}) \mbox{ for } t \in [0,1]  \mbox{ with }
 \bm{x}(1)=\bm{x}(0). 
\end{equation}
\review{\review{while} equation \eqref{linstab} becomes
\begin{equation} \label{linstabRes}
    \frac{d\bm{x'}}{dt}=T \bm{A}(\bm{t})\bm x'.
\end{equation}}
More specifically, eq. {\eqref{dynrescaled}} is a two-point BVP where periodic boundary conditions are imposed. 
This observation paves the way for alternate strategies to find PO owing to the numerous numerical techniques dedicated to two-point BVPs.\\

One of the most popular techniques for solving two-point BVPs is the shooting method and its variants, see e.g. \cite{atkinson1991introduction}. It is based on sequences of a repeated IVP solver, iteratively refining the initial value and the period to the point where periodicity is satisfied over the whole unit interval.
 The iterative process converges when the initial value lies on a stable or unstable periodic orbit and matches the end value within some prescribed accuracy. Overviews of shooting algorithms applied to
large multidimensional systems can be found in \citet{sanchez2004newton,viswanath2007recurrent,frantz2023krylov}. 
However, shooting methods may fail if the PO is too unstable. \review{Even if the initial value is close to a PO point, numerical integration can make a trajectory deviate rapidly from the desired PO. The perturbation to the PO can undergo such a
large amplification over a period that numerical convergence can be hampered \citep{duguet2008relative}.
}
\\

More robust alternate techniques to solve two-point BVPs can be used. The entire PO can be discretised as a finite set of points. The governing equations are satisfied at each of these points.
As an alternative, the PO
can also be expanded as a series on a well-chosen functional basis. The most straightforward option, given the temporal periodicity of the solution, is to consider a Fourier expansion, yet this is not the only possible choice. 
Instead of iterating only the initial condition and the value of the period as in iterative shooting methods, each iteration to reach the PO involves computing all the points of the (discretised) orbit -or all coefficients of the expansion- simultaneously. \review{The size of the linear system solved at each iteration scales linearly with 
$d$ times 
the number of points on the orbit or the number of retained coefficients.} Variants of these approaches combine highly accurate polynomial expansions between discrete points of the orbit. This technique is implemented for example in the software package AUTO \citep{Doedel_01}. \\

\review{In the approach retained here, we discretize the entire PO. We introduce  and compare three different discretizations to compute the periodic base solution 
$\bm x_p$ and predict its stability in the Floquet sense.}
\\



\subsection{Finite differences \review{discretisation}} \label{sec: fd}


Eq. \eqref{dynrescaled} is discretised at $N_P+1$ times corresponding to $N_P$ time intervals.  
In the simplest form the time derivative in Eq. \eqref{dynrescaled} is approximated using a second-order accurate Crank-Nicolson finite difference scheme~:
 \begin{equation} \label{fds}
 \dot{x}_i^{n+1/2}=\frac{x_i^{n+1}-x_i^{n}}{dt}
 \mbox{ for } n = 0,1,\dots,N_P-1  \mbox{ and } i =1,2, \dots, d 
 \end{equation} 
 where $i$ is the component of $
 \bm x$ expressed at the $n$th timestep, $n$ is the index of the temporal position on the orbit 
 and $dt$ is the time interval between two states $\bm x^n$ and $\bm x^{n+1}$. For simplicity, $dt$ is assumed constant equal to $1/N_P$. This leads 
 to a discrete set of equations :
  \begin{subequations}  \label{eq:FiniteDiffGlobal}
   \begin{equation}
       \frac{T}{2}  \left ( f(x_i^{n+1}) + f(x_i^{n}) \right ) - \frac{x_i^{n+1}-x_i^{n}}{dt}  = 0 \label{eq:FiniteDiffGlobala}
   \end{equation}
      \begin{equation}
          x_i^{N_P}-x_i^{0} = 0, \label{eq:FiniteDiffGlobalb}
      \end{equation}
      \begin{equation}
           \frac{x_1^{1}-x_1^{0}}{dt} = 0.  \label{eq:FiniteDiffGlobalc}
      \end{equation}
  \end{subequations}
 Equation \eqref{eq:FiniteDiffGlobala} is written for all $n$ 
 points $n=0,..,N_P-1$ on the orbit and all degrees of freedom $i=1,..,d$, resulting in  $N_p d$ scalar equations
 .
 Equation \eqref{eq:FiniteDiffGlobalb} imposes periodicity for $d$ degrees of freedom ($d$ scalar equations). 
Equation \eqref{eq:FiniteDiffGlobalc} closes the system for the extra unknown $T$ using a phase condition (one scalar equation). The exact number of equations and variables is therefore 
$(N_P +1)d + 1$. Instead of  \eqref{eq:FiniteDiffGlobalc}, other phase conditions can be considered (see chapter 7 in \cite{seydel2009practical}). Here, the time derivative of the first degree of freedom ($i=1$)
is imposed to be zero to second-order accuracy at the mid-point
between $n=0$ and $n=1$. The resulting algebraic nonlinear system~\eqref{eq:FiniteDiffGlobal} is solved by Newton-Raphson's method.\\

The Jacobian computed at each Newton-Raphson iteration is a square bordered block matrix of dimension $(N_P +1)d + 1$  of the form :
\begin{equation} \label{eq:JacobianNewton}
\left[ \begin{array}{cccccc|c}  
A_{0,0}  & A_{0,1}  & 0_{d} & \cdots & \cdots  & 0_{d} & \bm \alpha_0 \\
0_{d} & \hphantom{0}  A_{1,1}  & A_{1,2}  & \ddots &   & \vdots & \bm  \alpha_1 \\
\vdots & \hphantom{0} \ddots \hphantom{0} &  \hphantom{0} \ddots \hphantom{0} & \ddots & \ddots & \vdots & \vdots \\
\vdots & & \ddots & \ddots & \ddots & 0_{d} &  \vdots\\
0_{d} & \cdots & \cdots & 0_{d} & A_{N_P-1,N_P-1}  & A_{N_P-1,N_P}  & \bm \alpha_{N_P-1} \\
-I_{d}  & 0_{d} & \cdots & \cdots & 0_{d}  & I_{d}  & 0_{d1} \\
\hline
-\bm \beta  & \bm \beta & 0  & \cdots  & \cdots & 0  & 0  \\
\end{array} \right]
\end{equation}
where $I_d$ and $0_d$ are the identity and null $d \times d$ submatrices and $0_{d1}$ is the null column vector in $\mathbb{R}^d$. 
The submatrices $A_{n,n}$ $A_{n,n+1}$ have 
size  $d \times d$.
The submatrices  $A_{n,n}$  are obtained as the first-order partial derivative
of the left hand side of \eqref{eq:FiniteDiffGlobala} with respect to $x_i^n$ for all $i$ (with $n$ fixed). 
Similarly the submatrices $A_{n,n+1}$ are obtained by deriving the left hand side of \eqref{eq:FiniteDiffGlobala} with respect to $x_i^{n+1}$
 for all $i$ (with $n$ and $n+1$ fixed). 
The row vector ${\bm \beta} \in\mathbb{R}^d$ is  
given by $[1/dt,0,\cdots,0]$.
Finally, the
$\bm \alpha_n = 1/2 ( f(\bm x^{n}) + f(\bm x^{n+1}) )$ are column vectors of size $d$.

This specific structure of the Jacobian matrix is obtained by
choosing the 
ordering of the unknowns as in $[\bm x^0, \bm x^1, ..., \bm x^{N_P}, T]$. 
As the Jacobian matrix is sparse and bordered, efficient techniques  exist to solve the linear system
at each Newton step. 

\subsubsection*{Computation of the stability of the periodic orbit solution}

The stability of the periodic solution is found by introducing explicitly the complex Floquet multipliers $\sigma$
in an eigenvalue problem which  comes directly as a by-product of 
the computation of the periodic orbit. This approach was suggested by \cite{fairgrieve1991ok} in order to avoid forming the monodromy matrix. 
\review{Let us assume that the \review{periodic solution is encoded in a solution vector $\bm u=[\bm x^0, \bm x^1, ..., \bm x^{N_P}]$ and} is perturbed in the direction of one of the Floquet vectors $\bm v=[\bm x'^0, \bm x'^1, ..., \bm x'^{N_P}]$, itself associated with a Floquet multiplier $\sigma$.
%
Then the following constraint on ${\bm v}$ has to be satisfied :\begin{equation} \label{eq:FloquetMultiDef}
    { \bm x'^{N_P}} = \sigma{ \bm x'^0}.
\end{equation} 
}
The orbit is stable along the direction spanned by ${\bm v}$ if $|\sigma|<1$, neutrally stable if $|\sigma|=1$ and unstable if $|\sigma|>1$. 
In order to use the Jacobian matrix~\eqref{eq:JacobianNewton} in the eigenvalue problem with no extra computation, 
equation~\eqref{eq:FloquetMultiDef}, following \cite{fairgrieve1991ok},
is rewritten 
as~:
\begin{equation}
 {\bm x'^{N_P}} - { \bm x'^0} = \gamma \   {\bm x'^{N_P}}, \quad \mbox{ where } \gamma = 1 - \frac{1}{\sigma}.
\end{equation}
Linearisation of 
\eqref{eq:FiniteDiffGlobal} around the solution $(\bm u,T)$ leads to the generalised eigenvalue problem
\begin{equation}  \label{eq:generalizedEigValpb}
\bm{J} \bm{v} = \gamma \bm{B}\bm{v} . 
\end{equation}
The square matrix $\bm{J}$ of dimension $(N_P+1)d$ is extracted directly from the matrix \eqref{eq:JacobianNewton} by removing the last line and the last column. The square matrix $\bm{B}$ of dimension $(N_P+1)d$ is defined by :

\begin{equation} \label{eq:BmatrixFiniteDiff}
\left[ \begin{array}{ccccc|c} 
0_d & \cdots  & \cdots & \cdots  & 0_d & 0_d  \\
\vdots &  \ddots  & & & \vdots & \vdots  \\
\vdots & & \hphantom{0} 0_d \hphantom{0} & & \vdots & \vdots  \\
\vdots & & & \ddots   & \vdots & \vdots  \\
0_d & \cdots  & \cdots  & \cdots  & 0_d & 0_d  \\
\hline
 0_d & \cdots & \cdots & \cdots  & 0_d & I_d  \\
\end{array} \right].
\end{equation}


\review{This generalized eigenvalue problem involves the matrix $\bm B$, the very same one as considered by \cite{fairgrieve1991ok}.
Its block structure is shown in Eq. \ref{eq:BmatrixFiniteDiff} and contains as only non-null elements the $d \times d$ identity matrix\yd{,} denoted as $I_d$. In practice we used the routine \textit{eig()} of Matlab to find the $(N_P +1) \ d$ eigenvalues of the generalised eigenvalue problem~\eqref{eq:generalizedEigValpb}
if the size of the linear system is moderate. For larger systems, more specific
routines are used as detailed in section \ref{sec: dhc}.
Note that when the matrices $\bm J$ and $\bm B$ are real, the eigenvalues $\gamma$ and the eigenvectors $\bm v$ to \eqref{eq:generalizedEigValpb} are either real-valued or come in complex conjugate pairs.
$(N_P \ d)$ eigenvalues {, among the $(N_P +1) \ d$ eigenvalues output by the generalised eigensolver, are infinite. The relevant Floquet multipliers $\sigma_i=1,..,d$ are computed from the $d$ finite eigenvalues $\gamma_i=1,..,d$}. This is one of the main advantages of this method compared to, say, Hill's method (discussed in next section) : the number of meaningful eigenvalues is the one expected from theory{, namely $d$}. In particular there is no need to sort out the physically meaningful eigenvalues from spurious ones resulting from enlarging the problem from dimension $d$ to $(N_P+1)d$. Here we did not make any assumption on these $d$ eigenvalues (e.g. the existence of a neutral one as implied by the continuous dynamics, or the relation between them) and directly report them as output by the eigenvalue solver.}

 \subsection{\review{Fourier-Galerkin method}} 
 \label{sec:hbm}
A natural choice to cope with the periodicity of the PO is the use of trigonometric functions as a functional basis in a spectral expansion.
In Harmonic Balance Method, the periodic orbit is \review{
expanded} as a standard \review{truncated} Fourier series:
\begin{equation} \label{hbm}
\bm x(t) = \sum_{k=-N_F}^{N_F}\bm u_k e^{i 2 \pi k t},   \quad t \in [0, 1]
\end{equation}
where $N_F$ 
is finite
and the $\bm u_k$'s are complex coefficients.
When $\bm x(t)$ is real-valued, the property that, for all $k$, $\bm u_k$ and $\bm u_{-k}$ are complex conjugate, can be used to reduce the \review{number of real-valued degrees of freedom down to $2N_F+1$.} 
\review{As is common with spectral expansions, HBM allows for fast convergence provided the 
\review{target}
solution is smooth enough.} 
The complex coefficients $\bm u_k$ are found by introducing the decomposition \eqref{hbm} 
into the governing equations \eqref{dynrescaled}.
\review{
This yields the residual
\begin{equation}
    \bm r(t) \ { \equiv} \ T{\bm f} \left ( \sum_{k=-N_F}^{N_F}\bm u_k e^{i 2 \pi k t} \right ) - 2 \pi i \sum_{k=-N_F}^{N_F} k {\bm u_k } e^{i 2 \pi k t} 
\end{equation}
A set of $(2N_F+1)d$ algebraic equations for the ${\bm u}_k's$ is generated by requesting that the residual be orthogonal to each Fourier function.
We define the products
\begin{equation}
    \bm g_l=\int_0^1\bm r(t)e^{i2\pi l t}dt, \quad l=-N_F,..,N_F, 
\end{equation}
where the ${\bm g}_l$'s have dimension $d$,
and request that ${\bm g_l}=0,~l=-N_F,..,N_F$.}\\

The phase condition, similar to \eqref{eq:FiniteDiffGlobalc}, reads
\begin{equation}
    \dot{x}(0)=2\pi i \sum_{k=-N_F}^{N_F}k\ u_k=0.   
\end{equation}

Like for the finite differences method, these equations are solved using a standard Newton-Raphson method.
\review{The Jacobian matrix reads
\begin{equation} \label{jacHBM}
\left[
    \begin{array}{ccccc|c}
         \bm A_{-N_F,-N_F}& \quad\cdots\quad&\quad\cdots\quad&\quad\cdots\quad& \bm A_{-N_F,N_F} & \frac{\partial \bm g_{-N_F}}{\partial T}  \\
         \vdots&  \ddots &\ddots &\ddots & \vdots & \vdots  \\
         \vdots&  \ddots & \bm A_{l,k} &\ddots & \vdots & \frac{\partial \bm g_{l}}{\partial T}  \\
         \vdots&  \ddots & \ddots &\ddots & \vdots & \vdots  \\
         \bm A_{N_F,-N_F}&  \cdots&  \cdots&  \cdots & \bm A_{N_F,N_F} & \frac{\partial \bm g_{N_F}}{\partial T}  \\ \hline
         \bm \epsilon_{-N_F}&  \cdots&  \cdots&  \cdots & \bm \epsilon_{N_F} & 0
    \end{array}
    \right]
\end{equation}
where \begin{equation}
    \bm A_{l,k}=\frac{\partial\bm g_l}{\partial\bm u_k}
\end{equation}
and $\bm \epsilon_k$ is a row vector in $\mathbb{R}^d$ defined by
$${\bm \epsilon_k} = [ 2\pi i k,0,\cdots,0].$$
}
\review{Compared to the system arising from the finite difference discretisation, here} the Jacobian matrix is no longer sparse.

\subsubsection*{Computation of the stability of the periodic orbit solution}

In order to retrieve the stability of the orbit, Hill's method can be used. The details of the method are described by \cite{guillot2020purely} and we only summarize the main steps here.
\review{Let us assume again that the periodic solution ${\bm x}$ is perturbed in the direction of one
of the Floquet vectors ${\bm x'}$ itself associated with a Floquet
exponent $\mu$. The Floquet decomposition, 
together with the Fourier decomposition of the periodic function involved, leads in this particular case to :}

\begin{equation} \label{pert-form}
\bm x'(t)=\sum_{k=-N_F'}^{N_F'}\bm v_ke^{\mu t} e^{i 2 \pi k t }
\end{equation}
In practice it is 
assumed that $N_F=N_F'$ although this is no strict request.
The temporal growth or decay of the perturbation ${\bm x'}$ is governed by \review{
the exponential term $e^{\mu t}$ for the exponent $\mu$ with largest real part}.  
\review{As in the previous subsection, all the entries of ${\bm v}_k$ for $k=-N_F,...,N_F$ are stored in a vector ${\bm v}$ of size $(2N_F+1)d$.}
The expression \eqref{pert-form} is introduced into the linearised equation \eqref{linstabRes}. This yields the eigenvalue problem for $\mu$
\review{\begin{equation}
    \bm{J} \bm{v} = \mu \bm v,
    \label{jvmuv}
\end{equation}
where $\bm J $ is the Jacobian matrix \eqref{jacHBM} with the last column and last row removed.  Unlike the generalised eigenvalue problem \eqref{eq:generalizedEigValpb} arising in the finite difference case, it admits $(2N_F+1)d$ finite eigenvalues rather than only $d$.}\\

\review{One well-known problem associated with Hill's method is 
the difficulty of identifying which Floquet exponents are meaningful \citep{lazarus2010harmonic}}.  \review{As $N_F\to \infty$ the method is expected to output an infinity of identical Floquet multipliers that are identical modulo $2i\pi/T$. For finite truncation however, this exact degeneracy is lost
and one is left with a large set of $O(N_Fd)$ exponents to analyse. In particular the prediction of the exponent with largest real part can not be trusted \citep{kern2023linear}. This is a major limitation when predicting the stability of a given periodic orbit. } 
\\

A way out is to express the periodic orbit and especially the perturbation as a weighted sum of functions from another functional basis, one that is not by construction periodic. 
We consider in this study the 
basis of Chebyshev polynomials. 
Expanding a smooth enough function on a Chebyshev basis is known to minimise the pointwise error of the approximation of analytic functions \citep{boyd2001chebyshev}. 
\review{
This makes the Chebyshev basis a relevant choice over any other polynomial basis}. 

\subsection{Chebyshev polynomial expansion 
}
\label{sec:cheb}

We begin by expanding the periodic orbit solution as a finite sum of Chebyshev polynomials :
\begin{equation} \label{chebdec}
\bm x(\tilde{t})=\sum_{k=0}^{N_C}\bm a_k T_k(\tilde{t})=\sum_{k=0}^{N_C}\bm a_k \cos(k\arccos{\tilde{t}}),\ \ \tilde{t}\in [-1,1]
\end{equation}
where  the number of polynomials retained in the  truncation is $N_C +1 $. 
Since the polynomials $T_k(\tilde{t})$ are defined for $\tilde{t}\in [-1,1]$ a proper rescaling of the governing equations is required. 
Unlike Fourier trigonometric functions, the Chebyshev polynomials are not periodic by construction, hence the periodicity condition 
\review{for the PO}
\review{must now} be enforced.
The decomposition \eqref{chebdec} is introduced into the governing equation \eqref{dynrescaled} leading to :
\begin{equation} \label{eq:dynrescaledCheb}
{\bm f} \left ( \sum_{k=0}^{N_C}\bm a_k T_k(\tilde{t}) \right ) - \frac{2}{T} \sum_{k=1}^{N_C} k {\bm a_k } U_{k-1}(\tilde{t}) = 0,
\end{equation}
where the Chebyshev polynomials of the second kind $U_k(\tilde{t})$ are used to express the derivative of $T_k(\Tilde{t})$. The time derivative has to be transformed as $
    \frac{d \bm x}{d t }= 2 \frac{d \bm x}{d \Tilde{t}}$, 
where the original time $t\in[0,1]$  
is transformed into $\Tilde{t}\in[-1,1]$. \\


In order to generate a set of nonlinear algebraic equation~\eqref{eq:dynrescaledCheb} is evaluated at the 
{$[0,1,...,N_C-1]$}
Gauss-Lobatto points. 
We recall that Gauss-Lobatto points are
defined by
\begin{equation}
    \tilde{t}_l=-cos\left(\frac{ l \pi}{N_C}\right),\ \ \ l=0,1,2,...,N_C
\end{equation}
and span non-uniformly the whole interval $[-1,1]$. 
Writing equation~\eqref{eq:dynrescaledCheb} at the points $\tilde{t}_0 ... \tilde{t}_{N_C-1}$ yields $N_C \times d$ equations.
Enforcing the periodicity condition
\begin{equation}\label{ref:percond}
    \bm x(1)-\bm x(-1)=\sum_{k=0}^{N_C}\bm a_k(1-(-1)^k)=0
\end{equation}
using 
 $T_k(1)=1$ and $T_k(-1)=(-1)^k$ yields $d$ additional  equations.
The last equation comes from the phase condition : enforcing a zero derivative with respect to time 
at $\tilde{t} = -1$ to one specified degree of freedom (here $x_1$, the first component of $\bm x$) leads to~: 
\begin{equation}\label{ref:phasecondCheb}
     \frac{d x_1}{d t} |_{\tilde{t}=-1}\equiv \sum_{k=1}^{N_C} k \ {a_1}_k \ U_{k-1}(-1) =\sum_{k=1}^{N_C} {a_1}_k k^2 (-1)^{k+1} =0, 
\end{equation}
 where the ${a_1}_k$'s are the weights for $x_1$.
 The set of $(N_C+1)d + 1$ nonlinear algebraic equations is solved with the Newton-Raphson method.\\
 

The Jacobian matrix computed at each Newton-Raphson iteration is a square bordered block matrix of dimension $(N_C +1)d + 1$ of the following form,
assuming the following ordering of the unknowns $[\bm a_0, \bm a_1, ..., \bm a_{N_C}, 1/T]$
:
\begin{equation} \label{eq:JacobianNewtonChebyshev} 
\left[ \begin{array}{cccccc|c}  
A_{0,0}  & A_{0,1}  &  \cdots & \cdots & \cdots  & A_{0,N_C} & \bm \delta_0 \\
A_{1,0} & \hphantom{0}  A_{1,1}  & \cdots  & \cdots & \cdots & \vdots & \bm  \delta_1 \\
\vdots & \hphantom{0} \ddots \hphantom{0} &  \hphantom{0} \ddots \hphantom{0} & \ddots & \ddots & \vdots & \vdots \\
\vdots & \hphantom{0} \ddots \hphantom{0} &  \hphantom{0} \ddots \hphantom{0} & A_{l,k} & \ddots & \vdots & \bm \delta_l \\
\vdots & \hphantom{0} \ddots \hphantom{0} &  \hphantom{0} \ddots \hphantom{0} & \ddots & \ddots & \vdots & \vdots \\
A_{N_C-1,0} & \cdots & \cdots & A_{N_C-1,k} & 
\cdots & A_{N_C-1,N_C}  & \bm \delta_{N_C-1} \\
2  \ C_{d}^0  & \cdots & \cdots & 2 \ C_{d}^{k}  & \cdots & 2 \ C_{d}^{N_C}  & 0_{d1} \\
\hline
\bm \epsilon_0 & \bm \epsilon_1 &  \cdots & \bm \epsilon_{k}  & \cdots & \bm \epsilon_{N_C}  & 0  \\
\end{array} \right]
\end{equation}
$A_{l,k}$ and $C_{d}^{k}$  are $d \times d$ submatrices defined by :
$$
A_{l,k} = \frac{\partial {\bm f} \left ( \sum_{k'=0}^{N_C}\bm a_{k'} T_{k'}(\tilde{t}_{l}) \right )}{\partial \bm a_{k}} \mbox{ and }
C_{d}^{k} =   
\begin{cases}
    I_d & \text{ if $k$ is even} \\
    0_d & \text{ if $k$ is odd.} 
  \end{cases}
$$
The submatrices $I_d$ and  $0_d$ and the vector $0_{d1}$ are already defined in section~\ref{sec: fd}.
${\bm \delta_l}$ and $\bm \epsilon_k$ are respectively  column and row vectors in $\mathbb{R}^d$ defined by :
$$
{\bm \delta_l} = - 2 \sum_{k=1}^{N_C} k {\bm a_k } U_{k-1}(\tilde{t}_l) \mbox{ and }
{\bm \epsilon_k} = [  (-1)^{k+1} k^2,0,\cdots,0]. 
$$.
 
\subsubsection*{Computation of the stability of the periodic orbit solution} \label{stabCheb}
Once the spectral coefficients $\bm a_k$ of the periodic orbit and the period T are known, the linear stability of the orbit can be assessed. The procedure is closely related to that used for finite differences. We introduce a small perturbation to the solution
:
\begin{equation} \label{pertCheb}
\bm x'(\tilde{t}) =\sum_{k=0}^{N_C'}\bm a_k^\prime \ T_k(\tilde{t}) 
\end{equation}
where 
$N_C'=N_C$. 
For the perturbation we do not assume periodicity but rather 
\begin{equation}\label{chebratio}
    \bm x'(1)=\sigma \bm x'(-1)
\end{equation}
\review{where $\sigma\in\mathbb{C}$ is again interpreted as a Floquet multiplier controlling the stability of the periodic orbit. Inserting the condition \eqref{chebratio} into the ansatz \eqref{pertCheb} leads to :}
\begin{equation}
\sum_{k=0}^{N_C} \bm a_k^\prime T_k(1)=\sigma \sum_{k=0}^{N_C} \bm a_k^\prime T_k(-1).
\end{equation}
Eq. \eqref{chebratio} is decomposed as
as :
\begin{equation} \label{eq:FairgrieveTrick}
{\bm x'(1)} - { \bm x'(-1)} = \gamma \ {\bm x'(1)}  \quad \mbox{ where } \gamma = 1- \frac{1}{\sigma}.
\end{equation}
\review{The decomposition \eqref{pertCheb} is introduced into equation \eqref{linstabRes}.}
Evaluation 
at \review{the $N_C$ first} Gauss-Lobatto points yields the generalised eigenvalue problem 
\begin{equation} \label{chebevp}
\bm{J} \bm{a^\prime} = \gamma \bm{B}\bm{a^\prime}.
\end{equation}
Owing to equation~\eqref{eq:FairgrieveTrick}, the matrix $\bm J$ is directly determined by the Jacobian matrix \eqref{eq:JacobianNewtonChebyshev} computed at convergence (last iteration in Newton-Raphson solver) by removing the last row and the last column. Each eigenvector is ordered in the following way
${\bm a^\prime} = [\bm a_0^\prime, \bm a_1^\prime, ..., \bm a_{N_C}^\prime]$. The square matrix $\bm{B}$ of dimension $(N_C+1)d$ is defined by :
\begin{equation} \label{eq:BmatrixChebyshev}
{\bm B} = \left[ \begin{array}{ccccc|c} 
0_d & \cdots  & \cdots & \cdots  & 0_d & 0_d  \\
\vdots &  \ddots  & & & \vdots & \vdots  \\
\vdots & & \hphantom{0} 0_d \hphantom{0} & & \vdots & \vdots  \\
\vdots & & & \ddots   & \vdots & \vdots  \\
0_d & \cdots  & \cdots  & \cdots  & 0_d & 0_d  \\
\hline
 I_d & \cdots & \cdots & \cdots  & I_d & I_d  \\
\end{array} \right].
\end{equation}

The eigenvalue problem \eqref{chebevp}  has $(N_C +1) \ d$ eigenvalues. \review{Despite the different approximation to the solution compared to \cite{fairgrieve1991ok}, the $\bm B$ matrix involved in \eqref{chebevp} is the same as in that study. As a consequence, among the $(N_C +1) \ d$ eigenvalues computed using Matlab's \textit{eig()} routine (or specific routines for large linear systems),} $(N_C \ d)$ eigenvalues are infinite and the Floquet multipliers $\sigma$ are computed from the $d$ finite eigenvalues $\gamma$. The Floquet exponents $\mu_i=1,..,d$ are directly deduced from the Floquet multipliers $\sigma_i,~i=1,..,d$ via the relation $\sigma_i=e^{\mu_i T}$.
The periodic orbit is unstable if $\max{\vert \sigma_i\vert}  >1$ (i.e. $\max{(real(\mu_i))}>0$). The main property to exploit is that the eigenvalue problem \eqref{chebevp}  yields \textit{only} $d$ finite \review{multipliers that can directly be interpreted as numerical approximations of the true } Floquet multipliers. This is in contrast to Hill's method that yields $(2N_F+1)d$ finite exponents together with no obvious method to sort them out and select the meaningful ones.\\

\section{\review{Application to low-dimensional systems}} \label{examples}
In this section several examples of periodic orbits in autonomous dynamical systems are presented. 
The stability of the orbits, 
predicted by either real or complex Floquet multipliers, is computed successively using 
finite differences, 
 Fourier and Chebyshev spectral 
time discretisations.  

\subsection{Lorenz system \review{: three cases}}
The Lorenz system is governed by a system of three ordinary differential equations:

\begin{subequations} \label{lor1}
\begin{equation} 
\frac{dx}{dt^*}=s(y-x),
\end{equation}
\begin{equation}
\frac{dy}{dt^*}=x(\rho-z)-y,
\end{equation}
\begin{equation}
\frac{dz}{dt^*}=xy-\beta z
\end{equation}        
\end{subequations}
\review{\subsubsection{$\rho=24$}}

\begin{figure}
\centering



\includegraphics[width=0.49\textwidth]{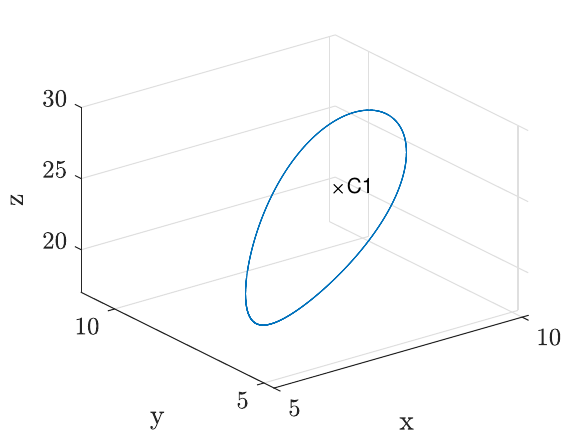}
\includegraphics[width=0.49\textwidth]{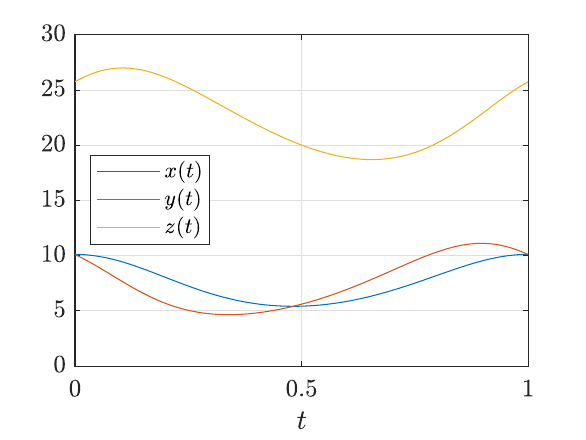}

\caption{
The unstable periodic orbit
($T = 0.6793$) 
for the Lorenz system $\rho=24$, $s=10$, $\beta=8/3$ as a state portrait on the left and in a component-wise representation over one period on the right. A stable fixed point is marked with $C_1=[\sqrt{\beta(\rho-1)},\sqrt{\beta(\rho-1)},\rho-1]=[8.76,  \   8.76,   \ 23]$. Representation obtained with HBM using $N_F=19$ modes. $\bm x(t=0)=(10.12,\ 10.1,\    25.76)$.
}
\label{fig:lor24}
\end{figure}

For $\rho=24$, $s=10$, $\beta=8/3$ the system admits 
an unstable periodic orbit born in a subcritical Hopf bifurcation at $\rho=24.74$ of a 
fixed point $C_1$ \citep{lorenz1963deterministic}, shown in figure \ref{fig:lor24}.  \\

\begin{figure}
    \centering
    \includegraphics[height=0.52\textwidth]{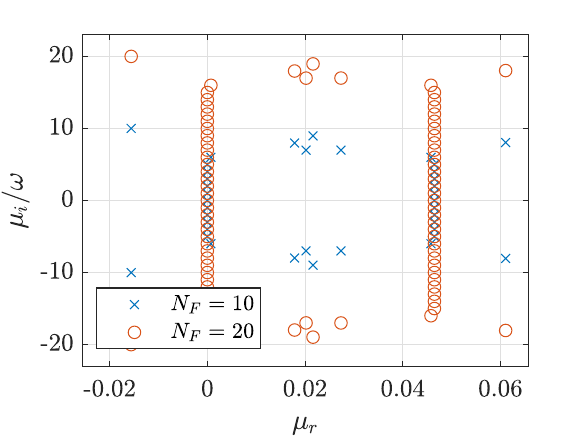}
    \includegraphics[height=0.52\textwidth]{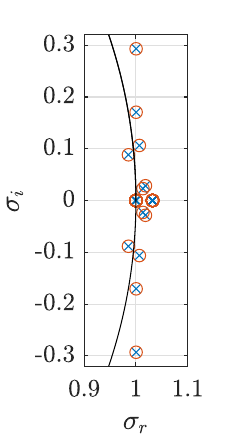}
    \caption{Left: Subset of Floquet exponents of the orbit of the
     Lorenz system for ($\rho=24$, $s=10$, $b=8/3$) found by Hill's method for $N_F=9$ and $N_F=19$. Only the Floquet exponents around $\mu_r=0.0$ and $\mu_r=0.0465$ are displayed. The region of the complex plane around $\mu_r=-13.713$ is omitted for clarity. Right: Corresponding Floquet multipliers $\exp(\mu T)$. Hill's method yields a set of exponents that have to be filtered to discard the spurious values. 
    }
    \label{fig:conv24}
\end{figure}

The stability of the orbit is found using each of the three methods presented in Section 2. 
The resulting spectrum and the convergence of the unstable Floquet exponent are shown in figure \ref{fig:conv24}. 
For the evaluation of the error on the exponents $\mu_3$ and $\mu_1$ 
we take as reference exponents ($\mu_\infty$) 
those extracted from Hill's matrix for $N_F=19$. The absolute error for each exponent is computed as $|\mu_{\infty}-\mu|$. 
The 
values of the three Floquet exponents are approximately $-13.713, 0.0, 0.0465$ for stable, neutral and unstable perturbations, respectively, with the value for the neutral exponent assumed exactly zero. 


First, as can be seen in figure \ref{fig:conv24}, the spectrum associated with Hill's matrix is populated by a large number of spurious eigenvalues, whereas for a three-dimensional ODE system such as the Lorenz system, only three eigenvalues are needed (one of them being necessarily neutral). 
In the absence of any strategy to filter the eigenvalues of Hill's matrix it can thus be challenging to draw conclusions concerning the orbit's stability. 

\begin{figure}
    \centering
    \includegraphics[width=0.32\textwidth]{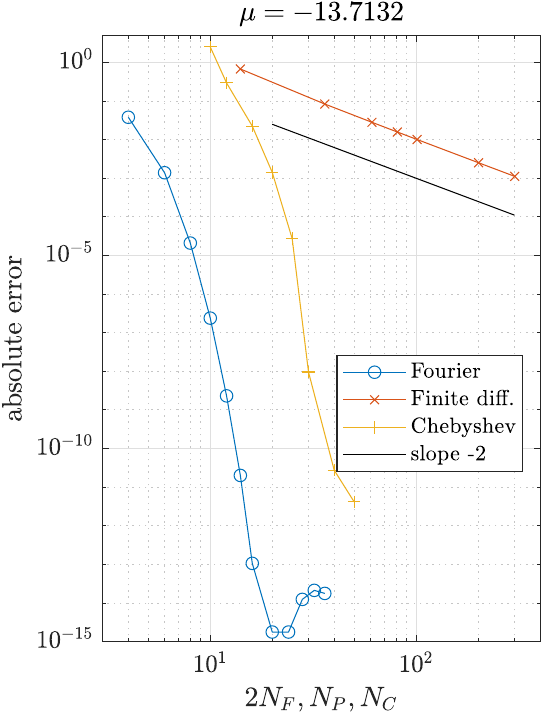}
    \includegraphics[width=0.32\textwidth]{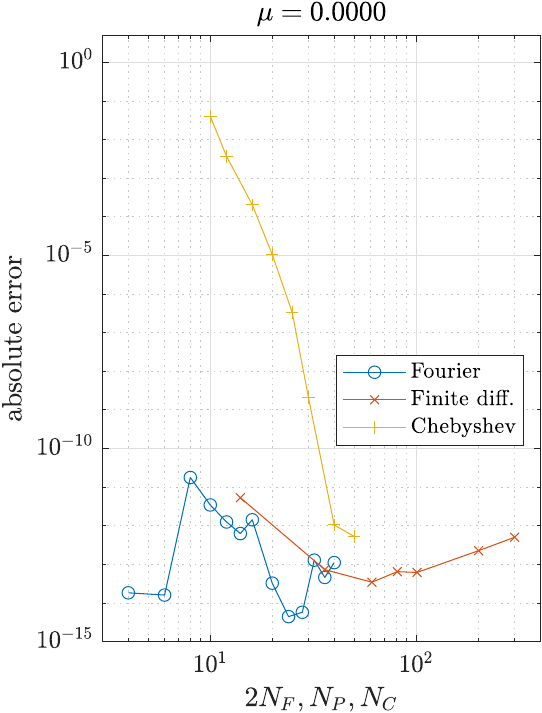}
    \includegraphics[width=0.32\textwidth]{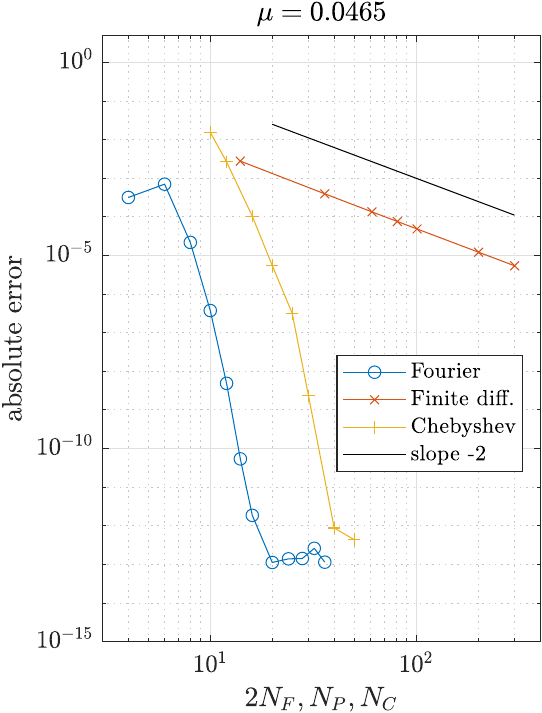}
    \caption{
    Error committed on each Floquet exponent for a periodic orbit of the Lorenz system for $\rho=24$, $s=10$, $b=8/3$
    computed with Finite Differences, Hill's method (Fourier) and Chebyshev method. Exponents are  displayed in increasing order from left to right. Hill's and Chebyshev methods show spectral convergence while Finite Differences only show second order convergence for the non-zero exponents. Results reported as a function of the number of real-valued degrees of freedom, namely $2N_F$, $N_P$ and $N_C$  respectively. 
    }
    \label{fig:three-conv}
\end{figure}

The convergence of the 
Floquet exponents is 
listed in table \ref{tab:convr24} and also 
illustrated graphically in figure \ref{fig:three-conv}. Convergence 
is reported for each exponent provided the resolution is high enough.
On one hand, both HBM and Chebyshev 
expansions are 
characterised by spectral exponential-like convergence, although 
slower for the Chebyshev method for the same number of degrees of freedom. On the other hand, as shown in figure \ref{fig:three-conv}, the finite differences method is only second order accurate in determining 
non-zero Floquet exponents. The second-order accuracy is 
consistent with
the fact that 
the approximation of the time derivative 
is also here second order. 
The advantage of Chebyshev expansions 
for identifying the 
relevant Floquet exponents is balanced by its  slower convergence compared to HBM. 
\\

 
\begin{table}[]
 \centering
\begin{tabular}{c|cccc}
     
$N_P$ & $\mu_3$& $\mu_2$& $\mu_1$ & $T$\\ \hline
13	&-14.3895880
&-3.82750361$\times10^{-12}$	&0.0437157935
&0.692268000
\\
35	&-13.7967351
&-2.90158308$\times10^{-14}$	&0.0460974317
&0.681075444
\\
60	&-13.7412922
&1.73082817$\times10^{-14}$	&0.0463577558
&
0.679926770
\\
100	&-13.7232514
&-1.51940104$\times10^{-13}$	&0.0464440859
&0.679548972
\\
300	&-13.7142788
&-9.08625324$\times10^{-14}$	&0.0464873253
&0.679360332
\\
\\
$N_F$ & $\mu_3$& $\mu_2$& $\mu_1$ & $T$\\ \hline
1	&-13.6751635
&1.84076530$\times10^{-14}$	&0.0468109907
&0.677821008
\\
4	&-13.7131592
&-3.46572972$\times10^{-12}$	&0.0464931094
&0.679336744
\\
9	&-13.7131594
&-3.27490584$\times10^{-14}$	&0.0464927338
&0.679336764
\\
19	&-13.7131594
&-1.11527694$\times10^{-13}$	&0.0464927338
&0.679336764
\\
\\
$N_C$ & $\mu_3$& $\mu_2$& $\mu_1$ & $T$\\ \hline
$11$& $-13.9789749 
$& $3.70357768 \times 10^{-3}$& $0.0438154932
$& $0.678437981 
$\\ 
$13$& $-14.1819050 
$& $2.67871156 \times 10^{-3}$& $0.0446843534
$& $0.679357954 
$\\ 
$21$& $-13.7134167 
$& $3.03491195 \times 10^{-6}$& $0.0464901602
$& $0.679336823 
$\\ 
$31$& $-13.7131593 
$& $-8.53469671 \times 10^{-10}$& $0.0464927345
$& $0.679336764
$\\ 
$51$& $-13.7131594
$& $-3.90264847 \times 10^{-13}$& $0.0464927338
$& $0.679336764
$\\ 


\end{tabular}
 \caption{
 Three Floquet exponents and the period $T$ of 
 one unstable periodic orbit of the Lorenz system for $\rho=24$, $s=10$, $b=8/3$. $N_P$, $N_F$ and $N_C$ 
 are the numerical resolutions corresponding respectively to Finite Differences, Hill's method and Chebyshev method.
The Fourier-Galerkin method reaches machine precision with only $N_F=9$ modes, while the Chebyshev method reaches three significant digits with $N_C=21$.
 }
 \label{tab:convr24}
\end{table}


\review{\subsubsection{$\rho=28$}}

\begin{figure}
 \centering
 \includegraphics[width=0.49\textwidth]{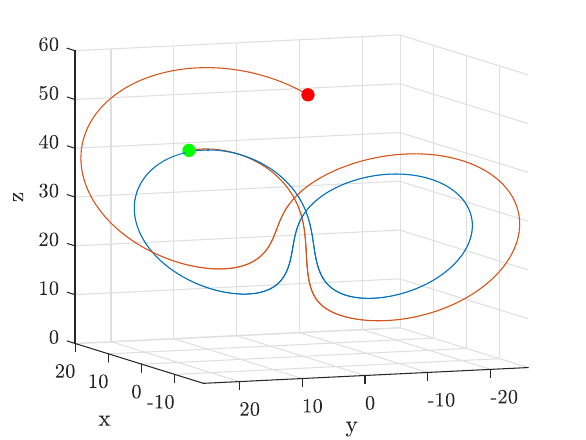}
 \includegraphics[width=0.49\textwidth]{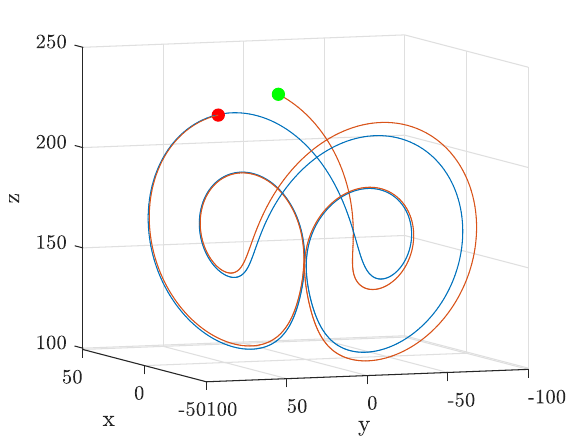}
 \caption{Periodic orbit solution of the
 Lorenz system for $\rho=28$ 
($T=1.559$)(left) and $\rho=160$ 
($T=1.153$)(right) in blue, together with an unstable (left panel) or stable (right panel) eigenvector (solid red line) computed with the Chebyshev method. 
Green dot : initial position, red dot : position after one period.
}
 \label{fig:lor28}
 \end{figure}

We report now on the convergence of the methods for more demanding parameter values. For $\rho=28$, $s=10$, $\beta=8/3$ the Lorenz system \eqref{lor1} 
admits a chaotic attractor and hence an infinite number of unstable periodic orbits
\citep{cvitanovic2005chaos}. 
The periodic orbit with the smallest period among those identified, labelled AB by \cite{viswanath2003symbolic}, is shown in figure \ref{fig:lor28}. For this specific orbit, the period found is $T=1.558652210716$ and the largest Lyapunov exponent is $\mu_1=0.9946$. 
\review{The remaining exponents are easily deduced from $\mu_1$. Since the tangent vector is a neutral direction, $\mu_2=0$. Using the property that the rate of growth rate of infinitesimal volumes $\nabla \cdot \bm f=-(s+1+\beta)=-13.666\bar{6}$ is independent of space and time \citep{lorenz1963deterministic}, we evaluate it at any point on the periodic orbit ${\bm x_p}$ where it is equal to the sum $\mu_1+\mu_2+\mu_3$. We deduce that $\mu_3=-14.661$. }
The three Floquet exponents 
\review{found numerically} are listed in table \ref{tab:lor28conv} \review{as a function of the temporal resolution used.}
\review{For the finite difference method (not included in the table),} the relative error
drops below 1\% for the period and 
\review{for all} Floquet exponents when $N_P \ge$ 80.\\



\review{\subsubsection{$\rho=160$}}

\begin{table}[]
\centering
\begin{tabular}{c|cccc}
$N_F$ & $\mu_3$& $\mu_2$& $\mu_1$ & $T$\\ \hline
11 &-14.661
&6.8400$\times10^{-15}$	&0.99542
& 1.55867856\\
15 &-14.661
&4.5433$\times10^{-15}$	&0.99466
& 1.55865243
\\
19 &-14.661
&-1.5859$\times10^{-14}$	&0.99465
& 1.55865221
 \\
\\

     $N_C$ & $\mu_3$& $\mu_2$& $\mu_1$  \\ \hline
$29$& $-5.691431$& $6.103783 \times 10^{-2}$& $0.9868981
$& $1.55723896$\\ 
$49$& $-9.451596$& $2.331945 \times 10^{-3}$& $0.9948799
$& $1.55863779$\\ 
$69$& $-12.28668
$& $5.785412 \times 10^{-5}$& $0.9946627
$& $1.55865221$\\ 
$89$& $-14.43058
$& $1.110621 \times 10^{-6}$& $0.9946504
$& $1.55865221$\\ 
$109$& $-14.65851
$& $1.734342 \times 10^{-8}$& $0.9946500
$& $1.55865221$\\ 

    \end{tabular}
\caption{Three Floquet exponents and period of an unstable periodic orbit of the Lorenz system for $\rho=28$, $s=10$, $b=8/3$ for HBM/Hill's ($N_F$) and Chebyshev ($N_C$) method. 
}
\label{tab:lor28conv}
\end{table}

\begin{table}[]
\centering
\begin{tabular}{c|cccc}
$N_F$ & $\mu_3$& $\mu_2$& $\mu_1$ & $T$\\ \hline
19 &-12.469
&-1.1952	&-7.7734$\times10^{-14}$	 & 1.15297839
\\
21 &-12.461
&-1.2055	&2.6118$\times10^{-15}$	& 1.15295801
\\
23 &-12.460
&-1.2060	&1.8593$\times10^{-14}$	  		& 1.15295091
\\
25 &-12.460
&-1.2067	&4.8119$\times10^{-14}$	& 1.15294951
\\
27 &-12.460
&-1.2068	&1.6459$\times10^{-15}$& 1.15294907
\\
\\
          $N_C$ & $\mu_3$& $\mu_2$& $\mu_1$ \\ \hline 
%
$49$& $-4.4525$& $-2.3052$& $-4.7402 \times 10^{-2}$& $1.15205890$\\ 
$69$& $-9.3502$& $-1.0985$& $9.1673 \times 10^{-3}$& $1.15284030$\\ 
$89$& $-11.272
$& $-1.1883$& $1.6520 \times 10^{-3}$& $1.15293896$\\ 
$109$& $-14.358
$& $-1.2044$& $2.2840 \times 10^{-4}$& $1.15294808$\\ 
$129$& $-12.609
$& $-1.2066$& $2.8140 \times 10^{-5}$& $1.15294886$\\ 
$149$& $-12.477
$& $-1.2069$& $3.2048 \times 10^{-6}$& $1.15294892$\\

\end{tabular}
\caption{Three Floquet exponents and the period of the stable periodic orbit of the Lorenz system for $\rho=160$, $\mu=10$, $b=8/3$ for HBM/Hill's ($N_F$) and Chebyshev ($N_C$) method.
}
\label{tab:lor160stab}
\end{table}

The last example 
from the Lorenz system is a stable periodic orbit. For $\rho\in(145,166)$ there is a periodic window \citep{strogatz2018nonlinear}. A stable orbit for $\rho=160$ is shown in figure \ref{fig:lor28}. Its 
Floquet exponents are listed in table \ref{tab:lor160stab}. 
As expected, for this stable orbit all non-zero exponents are negative apart from $\mu_1$ which is neutral. Hill's method with $N_F=23$ modes achieves four digits of accuracy for all Floquet exponents. Comparable accuracy on $\mu_2$ using the Chebyshev method requests, in comparison, as many as $N_C=129$ modes.\\

\review{\subsection{Lorenz system : convergence analysis}}

\begin{figure}
    \centering
    \includegraphics[width=\textwidth]{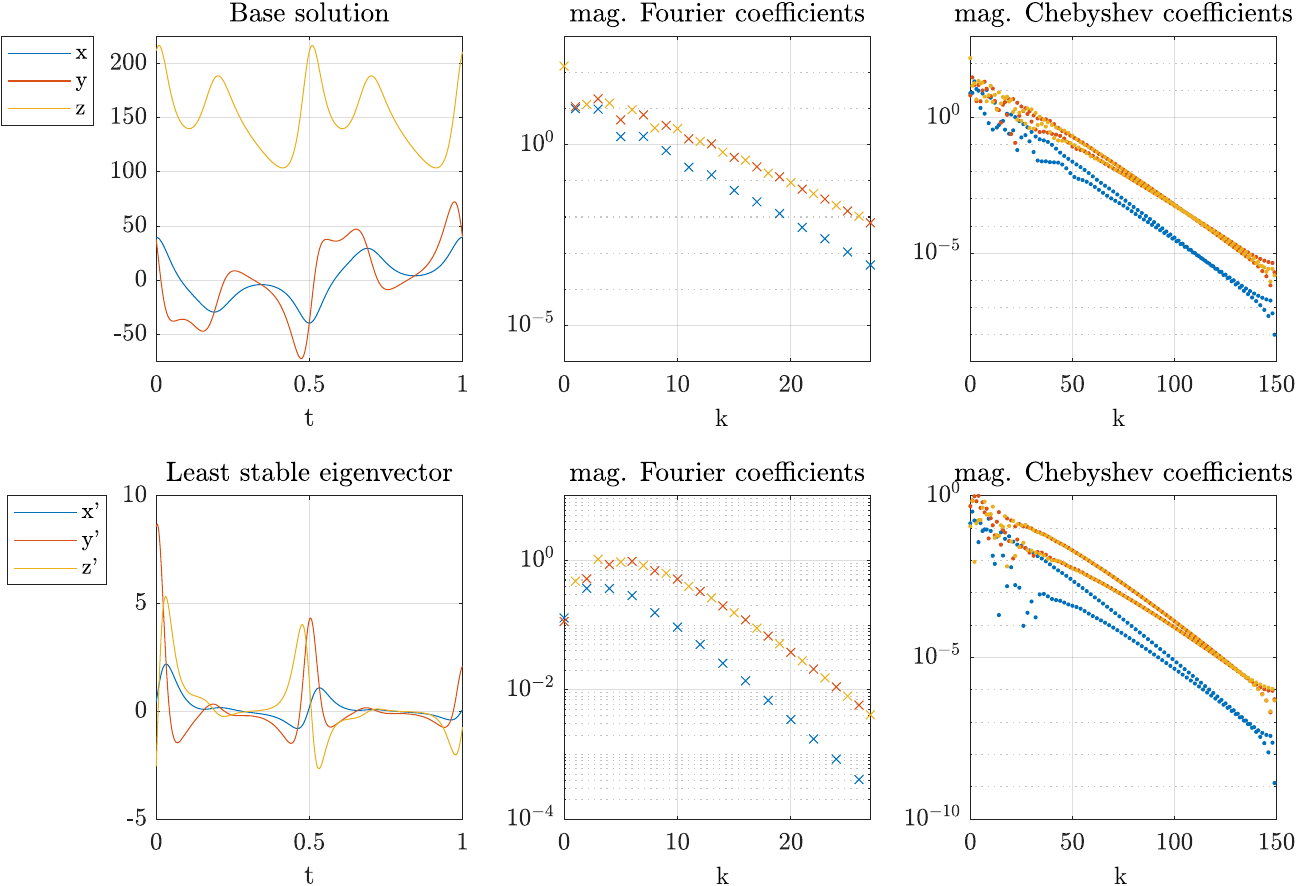}
    \caption{\review{Periodic orbit 
    solution (top) and least stable \review{Floquet} eigenvector (bottom) for the Lorenz system $\rho=160$, $s=10$, $b=8/3$ (cf. fig \ref{fig:lor28} right) and the corresponding magnitudes of Fourier and Chebyshev coefficients for cases $N_F=27$ and $N_C=149$. Whereas the base solution is periodic, its corresponding least stable eigenvector is not. Note that the Chebyshev coefficients fall to the eye on two distinct lines, which is related to a particular symmetry of the solution seen in the Fourier coefficients. Geometric convergence of the coefficients, marked by the constant slope in the lin-log plot, is characteristic for functions that are smooth in variable $t$. }
    }
    \label{fig:spectralcontent}
\end{figure}

\review{The convergence of finite differences and/or spectral methods is well documented in the literature. In particular finite differences are characterised by power-law convergence (i.e. the error decreases as inverse powers of the resolution $N_P$)
whereas for smooth enough solutions spectral methods, including Fourier and Chebyshev expansions, are characterised by (asymptotically faster) exponential convergence \citep{boyd2001chebyshev}.}\\

\review{
It is observed that, for equivalent accuracy, the number of Chebyshev coefficients is larger than the number of Fourier coefficients 
 for the cases corresponding to tables \ref{tab:lor28conv} and \ref{tab:lor160stab}. One reason for the slower convergence of the Chebyshev expansion could be the limited smoothness of the function being approximated. To investigate the convergence properties of the expansions \eqref{hbm} and \eqref{chebdec}, the magnitude of Fourier and Chebyshev coefficients corresponding to the base solution $\bm x_p$ and the least stable eigenvector $\bm x'(t)$ are plotted in figure \ref{fig:spectralcontent}. For both cases geometric convergence 
is observed. Following \cite{trefethen2019approximation}, the coefficients $a_k$ of the truncated Chebyshev expansion verify
\begin{equation}
    f(x)=\sum_{k=0}^{\infty} a_kT_k(x)=f_n(x)+\varepsilon_n
\end{equation}
with
\begin{equation}
    f_n(x)=\sum_{k=0}^{n} a_kT_k(x),
\end{equation}
and the difference between the truncated expansion and the function $|\varepsilon_n|=||f-f_n||_{\infty}$ both converge geometrically according to
\begin{equation}
    |a_k|\leq\frac{2M}{\rho^k}
\end{equation}
\begin{equation}
    ||f-f_n||_{\infty}\leq\frac{2M\rho^{-n}}{\rho^{-1}}
\end{equation}
provided $f$ is analytic in a Bernstein ellipse $E_{\rho}$ around the interval $x\in[-1,1]$ with $|f(x)|<M$. Furthermore,  the following inequality holds \citep{sinha1991efficient} for large enough $n$: 
\begin{equation}
|\varepsilon_n| \le |a_{n}|
\label{eq:sinhawu2}
\end{equation}
which relates the $L_\infty$ approximation error to the last coefficient of the computed expansion.\\
Therefore, since 
$\bm x$ and $\bm x'$ are at least analytic in $t\in[0,1]$ 
the large number of coefficients needed in practice is caused by the rich dynamics of the periodic orbit and its eigenmodes rather than by the 
lack of regularity
of the solution.\\}


\review{
The speed of convergence of the Fourier expansion with respect to the Chebyshev expansion can be explained by looking at the particular shape of the eigenvectors in figure \ref{fig:spectralcontent}. The eigenvector shows a narrow bump in the middle of the interval, which is known to hamper the speed of convergence of the Chebyshev expansion by comparison with the Fourier expansion.  The ratio $N_C/N_F$ observed in Table \ref{tab:lor160stab} to reach the same accuracy is consistent with the rule-of-thumb $N_C/N_F=3.5$ put forward by
\cite{boyd2001chebyshev}.\\}

\subsection{Langford system
}
\begin{figure}
\centering
\includegraphics[width=0.49\textwidth]{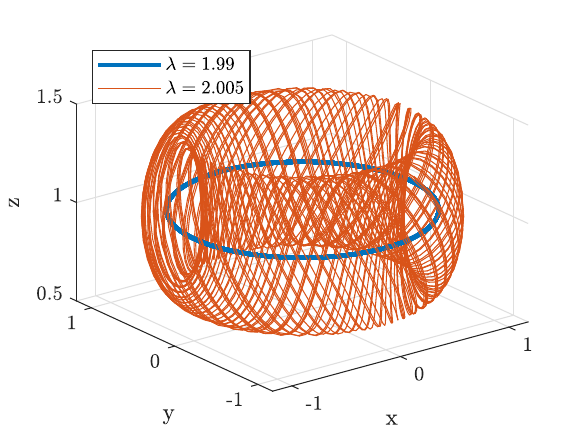}
\includegraphics[width=0.49\textwidth]{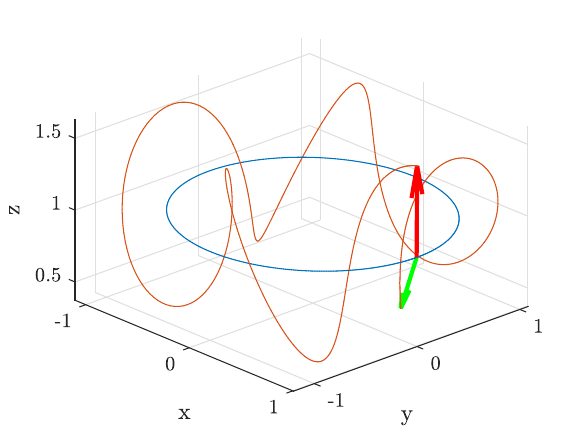}
\caption{Left: trajectory of the Langford system obtained using a time integrator for values of $\lambda$ below (blue) and above $\lambda_c=2$ (red). For $\lambda=1.99$ the trajectory converges asymptotically to the stable perfectly circular periodic orbit. For $\lambda=2.005$ the trajectory diverges towards a stable torus. Right: the unstable periodic orbit and its eigenvector for $\lambda=2.005$, obtained using the Chebyshev method.
$\bm x'(t=0)$ is marked with a green and $\bm x'(t=1)$ with a red arrow.}
\label{fig:lang}
\end{figure}

The periodic orbits of the Lorenz system presented so far 
were either stable or unstable, but always with \textit{real} 
Floquet multipliers. The question arises whether the Chebyshev expansion 
can also handle the analysis of periodic orbits, stable or unstable,  characterised by \textit{complex} Floquet multipliers. Whenever 
stability is lost due to a pair of
non real
Floquet multipliers crossing the unit circle in the complex plane, a Neimark-Sacker bifurcation occurs, leading to additional oscillations on top of the PO. 
A low-dimensional example of such dynamics is the three-dimensional Langford ODE system \citep{seydel2009practical} given by 
\begin{subequations}
    \begin{equation}
 \dot{x}=(\lambda-3)x-0.25y+x\left(z+0.2\left(1-z^2\right)\right)
\end{equation}
\begin{equation}
\dot{y}=(\lambda-3)y+0.25x+y\left(z+0.2\left(1-z^2\right)\right)
\end{equation}
\begin{equation}
\dot{z}=\lambda z-\left(x^2+y^2+z^2\right)
\end{equation}
\end{subequations}
The system is parametrised by the real parameter $\lambda$. For $\lambda>1.683$,
it possesses an exactly circular periodic orbit whose analytical expression is given by
\begin{subequations}
\begin{equation}
    x=r\cos(0.25 t)
\end{equation}
\begin{equation}
    y=r\sin(0.25 t)
\end{equation}
\begin{equation}
z=2.5\left(1-\sqrt{0.8\lambda-1.24}\right)
\end{equation}
\end{subequations}
with $ r=\sqrt{z(\lambda-z)}$.
This PO
loses stability 
above $\lambda=2$. Due to its
simple analytical expression 
it is possible to express analytically its stability exponents as:
\begin{equation} \label{langEx}
\mu=\frac{\lambda-2z+\sqrt{(\lambda-2z)^2-8r(r-0.4rz)}}{2}
\end{equation}
where $z$ 
denotes the position of the $xy$ plane containing the orbit (see figure \ref{fig:lang}), $r$ 
its radius and $\mu$ to the complex Floquet exponent of the orbit. 
At $\lambda=2$ ($z=1$, $r=1$) the orbit loses stability to a complex pair of Floquet exponents $\mu=\pm i\frac{\sqrt{30}}{5}$ and 
leaves place to a $2$-torus. The periodic orbit, the torus and the instability mode computed using the Chebyshev method are all shown in figure \ref{fig:lang}. \review{
A pair of unstable non-trivial Floquet exponents is reported in table \ref{tab:lang} depending on the resolution.} Using $N_C=29$ modes six digits of the real part of the unstable Floquet exponent 
are computed accurately. The Chebyshev expansion is hence able to capture the oscillatory bifurcation.


\begin{table}[]
\centering

\begin{tabular}{c|ccc}
$N_C$ & $\mu_{2,3}$ ($\lambda=1.995$)& $\mu_{2,3}$ ($\lambda=2.0$)& $\mu_{2,3}$ ($\lambda=2.005$) \\ \hline
17 &$0.083759
\pm 1.0749i$&$1.7899\times 10^{-2}\pm 1.0780i$&$0.027143
\pm 1.0810i$ \\
21 &$-0.010471
\pm 1.0901i$&$5.0099\times 10^{-4} \pm 1.0958i$& $0.011387
\pm 1.1014i$ \\
25 &$-0.010858
\pm 1.0895i$&$1.4087\times 10^{-6} \pm 1.0954i$&$0.010815
\pm 1.1011i$\\
29 &$-0.010856665
\pm 1.0895400i$&$-3.9801709\times10^{-9} \pm 1.0954452i$ & $0.010810355
\pm 1.1011046i$ \\ \hline
exact& $-0.010856610 
\pm 1.0895398i$&$\pm 1.0954451i$&$ 0.010810312
\pm 1.1011045i$ \\
\end{tabular}
\caption{\review{}{Non-trivial Floquet exponents for the Langford system.}
The row labelled \textit{exact} corresponds to the analytical value of the exponent calculated using \eqref{langEx}. }
\label{tab:lang}
\end{table}
\section{Differentially Heated Cavity 
} \label{sec: dhc}
In order to check the applicability of the Chebyshev expansion in time to a 
much higher-dimensional problem, we consider the buoyancy-driven flow inside an air-filled differentially heated cavity with perfectly conducting horizontal top and bottom boundaries. This configuration was shown experimentally to undergo a transition to unsteadiness for values of the Rayleigh number $Ra =\frac{g \beta \Delta T H^3}{\nu \alpha} $ around $\approx 3\times10^6$~(\cite{briggs1985two}). It was
much used in the late 
1980's as a test case 
to predict numerically the transition to unsteadiness. Different numerical algorithms 
have 
provided the 
critical value 
for the loss of stability of the steady solution at $Ra_c \approx2.108\times10^6$ (\cite{winters1987hopf, Gelfgat&Tanasawa_1994, xin_plq_01}). 
Numerical simulations of the unsteady equations, reported in \cite{le1994onset}, showed that the nonlinear unsteady solutions above $Ra_c$ belong to different branches of solutions according to their oscillation period (see figure 5 therein) and that the first branch of time-periodic solutions born at $Ra_c$ loses its stability
 at a value of $Ra$ around $7\times10^6$. This flow configuration thus constitutes a good candidate to test the applicability of algorithms aimed at computing the instability of periodic solutions of the Navier-Stokes equations.

\subsection{Governing equations and numerical methods
}
The non-dimensional velocity $\mathbf{u}=(u_x,u_y)$, pressure $p$ and temperature $\theta$ obey the Navier-Stokes equations under the Boussinesq approximation:

        \begin{subequations}
        \begin{equation}
\frac{\partial \bm{u}}{\partial t}+( \bm{u}\cdot \nabla) \bm{u}=-\bm{\nabla} p + \frac{Pr}{\sqrt{Ra}}\bm{\nabla}^2  \bm{u}+Pr \theta \bm{e}_y 
\end{equation}
        \begin{equation}
\frac{\partial \theta}{\partial t}+( \bm{u}\cdot \nabla) \theta= \frac{1}{\sqrt{Ra}}\bm{\nabla}^2  \theta 
\end{equation}
\begin{equation}
 \bm{\nabla} \cdot  \bm{u}=0
\end{equation}
\end{subequations}
with a Prandtl number $Pr=0.71$. The square cavity is shown in figure \ref{fig:base22} together with the boundary conditions. For the sake of brevity we refer to~\cite{xin_plq_01} for a more detailed presentation of the physical and dimensionless parameters. 


The cavity is discretised with a Finite Volume method using $N_x\times N_y=128\times128$ cells resulting in $O(4N_xN_y)\approx65\ 000$ degrees of freedom (see \cite{Faugaret_jfm_2022}). 
To account for the existence of thin boundary layers along the walls, non-uniform grids in $x$ and $y$ are used, the cell boundaries being distributed  as :
\begin{equation}
    x_i=0.5 \:( 1 -  \cos(i \pi /N_x) )\quad i=0,1,...,N_x,
\end{equation}
\begin{equation}
    y_j=0.5 \:(1 - \cos(j \pi /N_y))\quad j=0,1,...,N_y.
\end{equation}
For the given resolution, the steady base flow is 
converged to machine precision using a Newton-Raphson method. The base flow exhibits a symmetry 
with respect to the cavity center that reads~:
\begin{equation} \label{sym}
    S:\ (x,y)\rightarrow (1-x,1-y),\quad (u_x,u_y,\theta)\rightarrow(-u_x,-u_y,-\theta).
\end{equation}
while perturbations can be invariant under $S$ or under 
\begin{equation} \label{sym2}
    S_2:\ (x,y)\rightarrow (1-x,1-y),\quad (u_x,u_y,\theta)\rightarrow(u_x,u_y,\theta).
\end{equation}

Its linear stability is computed by forming an eigenvalue problem and solving it numerically using ARPACK. More details can be found in \cite{gesla2023subcritical}.

\subsection{Instability of the 
Hopf branch}


\begin{figure}
    \centering
    \includegraphics[width=0.69\textwidth]{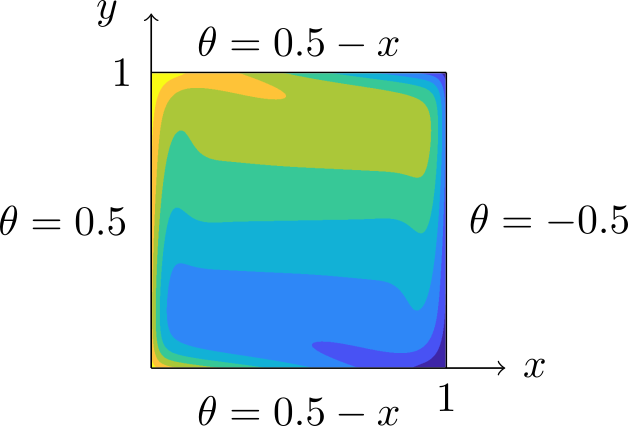}
    \caption{Square differentially heated cavity with perfectly conducting top and bottom walls. Temperature $\theta$ field of the steady base flow at $Ra=2.2\times10^6$ together with the boundary conditions. No slip conditions is assumed at the walls. Gravity points downwards towards decreasing $y$. The colormap spans the interval (-0.5,0.5) uniformly with 8 distinct colors. 
    \review{The solution verifies the symmetry  
    $S$. 
    } }
    \label{fig:base22}
\end{figure} 
The temperature field for the 
steady base flow is shown in figure~\ref{fig:base22} for $Ra=2.2\times10^6$. At $Ra_c\approx 2.1\times 10^6$ 
it loses its stability 
in a supercritical Hopf bifurcation that gives rise to a stable periodic orbit for $Ra>Ra_c$. 
To trace out the amplitude of the cycle born at $Ra_{c}$, an observable linked to  vorticity is monitored. We use the $L_2$-norm of the 
vorticity perturbation \review{to define
\begin{equation} \label{obser}
          A=r.m.s.\left({\sqrt{\int \vert\omega -\omega_{b}\vert^2 dx dy}}\right),
    \end{equation} 
where $\omega=\partial_x u_y-\partial_y u_x$ is the 
vorticity, $\omega_b$ is the equivalent quantity for the base flow, and the r.m.s. is applied over time.}
\begin{figure}
    \centering
\includegraphics[width=0.6\textwidth]{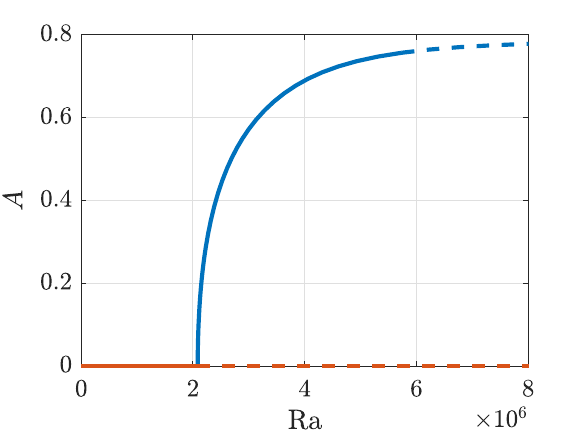}
    \caption{
    Bifurcation diagram for a square differentially heated cavity for the observable $A$ defined in \eqref{obser}. 
    The steady base flow is stable for $Ra_c<2.108\times10^6$ (red solid line) and unstable above $Ra_c$ (red dashed line). The supercritical Hopf branch of periodic orbits emanating from $Ra_c$ is stable for $Ra_{NS}<6.1\times10^6$ (blue solid line) and unstable above $Ra_{NC}$ (blue dashed line). 
    }
    \label{fig:bif-diag-dhc}
\end{figure}
A bifurcation diagram is shown in figure \ref{fig:bif-diag-dhc} 
$A$
as a function of $Ra$. \review{At $Ra_c$ the branch of periodic solutions is born. The periodic orbit is described at each subsequent \review{value of} $Ra$ with HBM. Continuation in $Ra$ is performed using standard pseudo-arclength continuation method \citep{allgower2012numerical}. Stability of the periodic solution can be 
investigated with Hill's method in \review{the} case of a Fourier expansion in time or with Chebyshev \review{expansion}, as proposed in this work.}

\review{The evolution of the most unstable Floquet exponent
of the supercritical branch is investigated using the Chebyshev expansion in time with $N_C=25$ 
and is listed in table \ref{tab:dhcstab}.}
It shows that the periodic orbit loses its stability to a pair of complex conjugate Floquet exponents in a Neimark-Sacker bifurcation
slightly above $Ra=6\times10^6$, thus confirming the result by \cite{le1994onset}.  

 \begin{table}[]
\centering
\begin{tabular}{c|cc}
Ra & $\mu$ & $\vert e^{\mu T }\vert$ \\ \hline
$5.0\times10^6$ & $-3.17\times10^{-3}\pm2.33\times10^{-1}i$ 	&9.855$\times10^{-1}$ \\
$5.5\times10^6$ & $-1.39\times10^{-3}\pm2.32\times10^{-1}i$ 	&9.936$\times10^{-1}$ \\
$6.0\times10^6$ & $-1.84\times10^{-4}\pm2.32\times10^{-1}i$ 	&9.991$\times10^{-1}$ \\
$6.5\times10^6$ & $5.64\times10^{-4}\pm2.32\times10^{-1}i$ 	&1.003 \\
$7.0\times10^6$ & $9.52\times10^{-4}\pm2.32\times10^{-1}i$ 	&1.004 
\\
\end{tabular}
\caption{Evolution of the leading Floquet exponent and multiplier of the periodic branch in the differentially heated cavity computed with the Chebyshev method ($N_C = 25$). The loss of stability occurs for $Ra$ 
between $6\times10^6$ and  $6.5\times10^6$. 
}
\label{tab:dhcstab}
\end{table}

\begin{figure}
    \centering
    \includegraphics[width=0.49\linewidth]{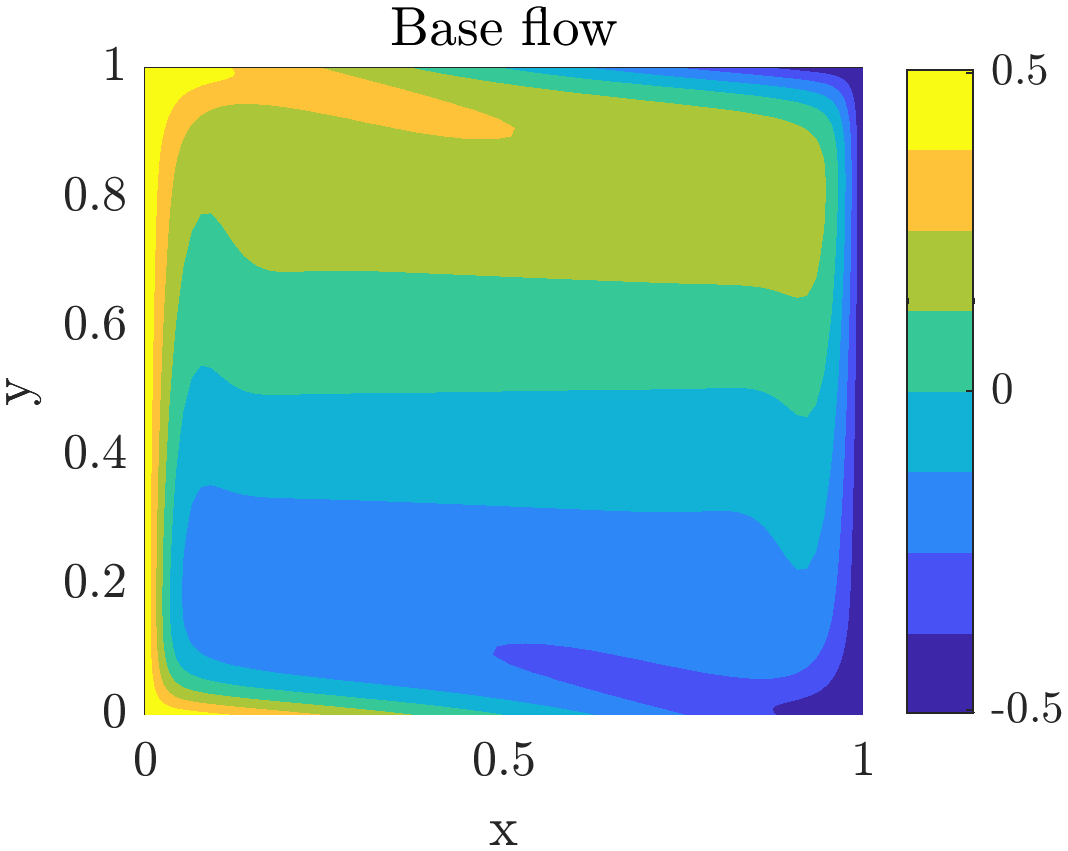}
    \includegraphics[width=0.49\linewidth]{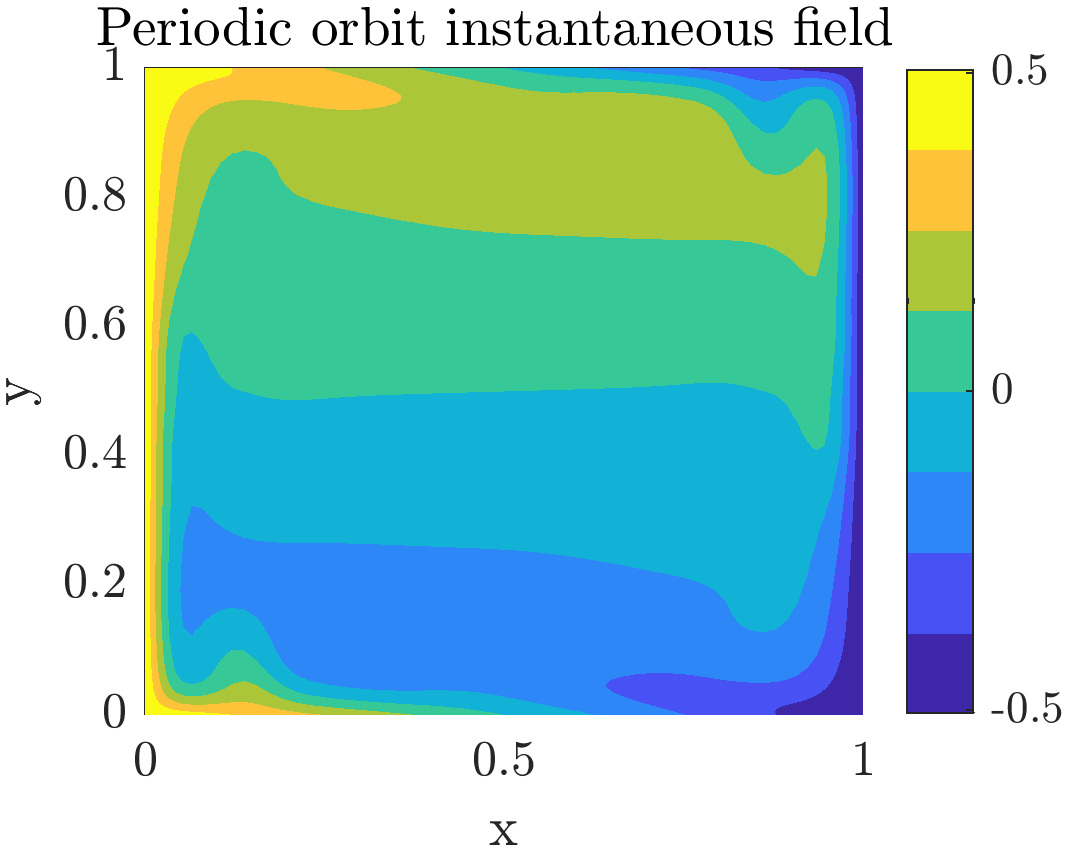}
    \includegraphics[width=0.49\linewidth]{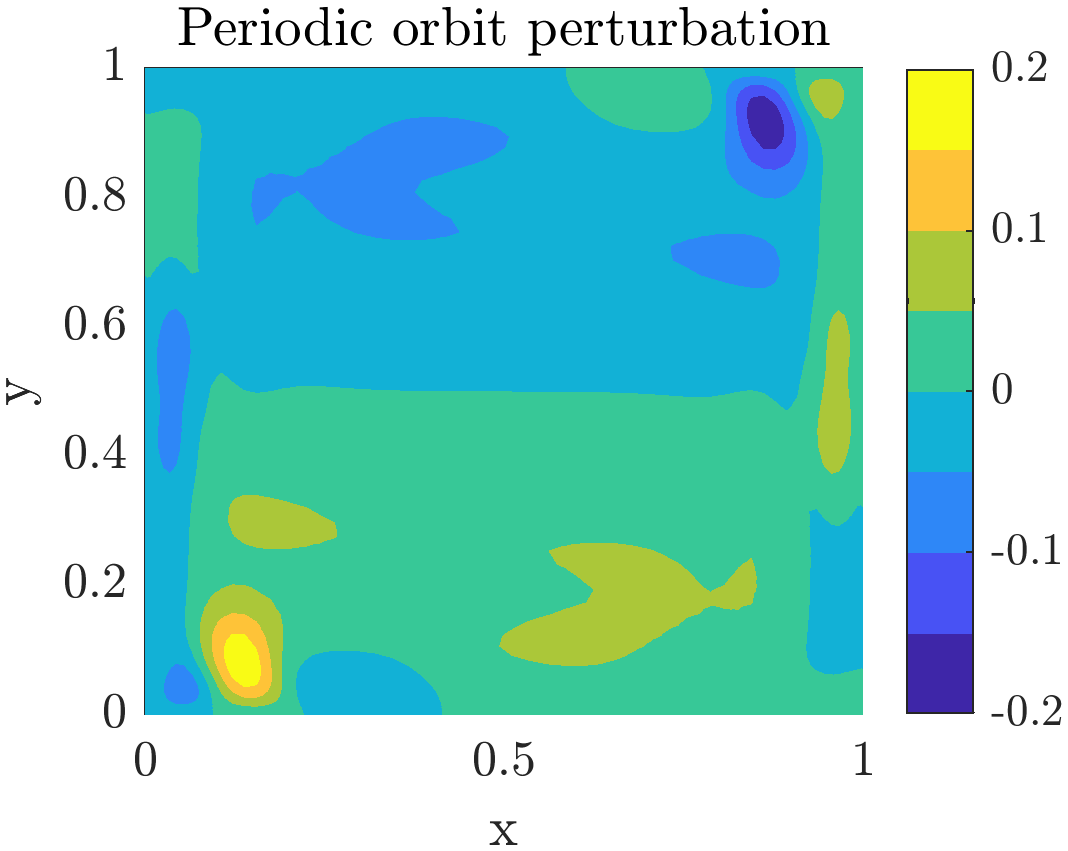}
    \includegraphics[width=0.49\linewidth]{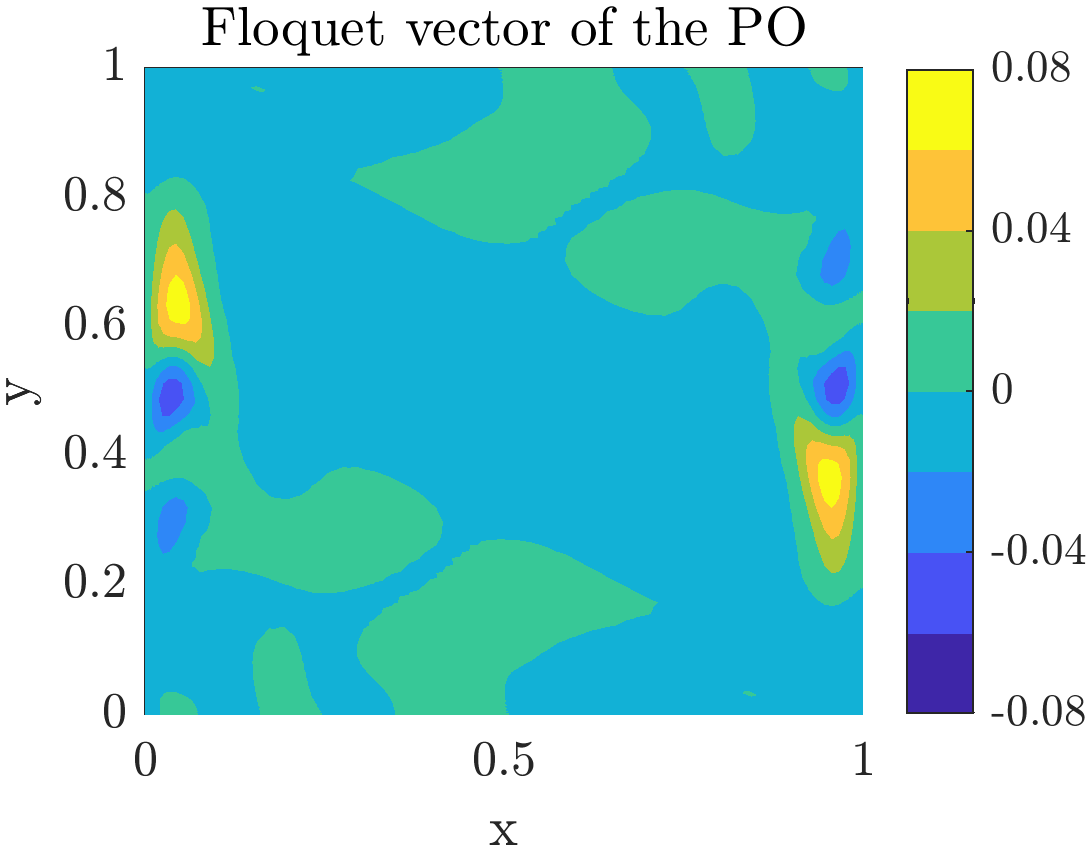}
    
    \caption{Temperature contours of the base flow (top left), instantaneous field of \review{the} saturated periodic orbit (top right), instantaneous field of saturated perturbation of the periodic orbit (bottom left) and its least stable Floquet vector (bottom right) for $Ra=5\times10^6$. Note that the base flow and the saturated periodic orbit temperature fields obey the symmetry 
    $S$, while the Floquet vector \review{
    obeys the other symmetry $S_2$.
    } }
    \label{fig:3panelcomp}
\end{figure}

Table~\ref{tab:hbmcheb} provides a comparison of the convergence of the largest Floquet exponent as a function of the number of temporal degrees of freedom 
in Fourier or Chebyshev expansion.
For the present spatial discretisation at $Ra=5\times10^6$, two correct digits of the real part of the leading Floquet exponent and five correct digits of the period 
require
$N_C=29$ Chebyshev modes. The linearly unstable base flow is shown in figure \ref{fig:3panelcomp}\review{, together with \review{temperature perturbation} snapshots \review{
along} the first branch and its least stable \review{Floquet} eigenvector}.  
\review{As previously observed in section \ref{examples}, the Fourier expansion converges faster than the Chebyshev expansion. 
\review{The sorting issue of Hill's method was avoided by first computing approximate values of the Floquet multipliers with the Chebyshev algorithm.  }
\review{In other words, t}aking advantage of the fast convergence properties of Fourier expansions is only possible once the correct eigenmodes have been \review{unambiguously} identified.
\review{This shows that both methods can be used in a complementary way.}}

\begin{table}[]
    \centering
    \begin{tabular}{c|ccccc}
         & $\mu$  & $T$ \\ \hline
         $N_F=4$&  $ -2.2585\times10^{-3} \pm 2.3229\times10^{-1}i$ & 4.60294436 \\ 
         $N_F=7$& $-3.1232\times10^{-3} \pm 2.3275\times10^{-1}i$ & 4.60715822 \\ 
         $N_F=9$& 
  $-3.1425\times10^{-3} \pm 2.3279\times10^{-1}i$ & 4.60701887 \\ 
  
  $N_F=11$&    $-3.144160\times 10^{-3}\pm 2.3278427\times10^{-1}i$ &4.60700620& \\ 
  $N_F=14$&   $-3.144645\times 10^{-3}\pm2.3278433\times10^{-1}i$    & 4.60700489\\ \hline 
  $N_C=19$&$-2.990\times10^{-3}\pm2.3257\times10^{-1}i$&4.60714417\\ 
         $N_C=29 $&$-3.156\times10^{-3}\pm2.3280\times10^{-1}i$ & 4.60701275\\ 
         $N_C=39$&$-3.14436\times10^{-3}\pm2.3278422\times10^{-1}i$& 4.60700519\\ 
         $N_C=49$ &$-3.14469\times10^{-3}\pm2.3278436\times10^{-1}i$&4.60700485& \\
    \end{tabular}
    \caption{
    Convergence study for the least stable Floquet exponent $\mu$ and the period $T$ of the stable orbit of the differentially heated cavity obtained with 
    Fourier method ($N_F$) and Chebyshev method ($N_C$). $Ra=5\times10^6$. 
    }
    \label{tab:hbmcheb}
\end{table}

\section{Summary} \label{summary}

\review{We have presented and compared three methods
to compute periodic solutions of multi-dimensional systems of ordinary differential equations
and to investigate
their stability characteristics. Three different time discretisations investigated include finite differences and spectral discretisation using Fourier modes (also known as Harmonic Balance method) or Chebyshev polynomials. All three methods rely on computing the periodic orbit using Newton-Raphson iterations. The linear stability of the periodic solution is investigated by forming a generalised eigenvalue problem without the need to consider, to form let alone to store a monodromy matrix. When the PO obeys to a PDE involving spatial and time variables, the observation that the time dependency is recast into a two-point BVP is of practical interest. Time-dependence can be handled as any spatial variable and thus, the implementation of the three presented methods in an existing PDE steady-state solver requires no conceptual change.}


\review{While the finite-difference time discretisation is only algebraically accurate, both Fourier-Galerkin and Chebyshev methods share similar spectral convergence characteristics. In practice however, Fourier-Galerkin requires less modes to reach a given accuracy. The true advantage of the Chebyshev method is brought to the forefront when extracting the stability of the orbit.
While the HBM is efficient to compute the periodic orbit, when coupled to Hill's method, it is known to yield a spectrum  with non-trivial sorting issues which are still the subject of ongoing research.
The finite difference and Chebyshev methods
output a directly usable spectrum, thereby avoiding the sorting issues associated with Hill's method.
This property is directly associated with the structure of the $\bm B$ matrix appearing in the generalised eigenvalue problem governing the stability of the periodic orbit.
The three methods were tested on selected stable or unstable periodic orbits of the Lorenz system, on a Neimark-Sacker bifurcation in the Langford system, as well as on a two-dimensional cavity flow governed by the incompressible
Navier-Stokes equations. This demonstrates the potential of Chebyshev method in conjunction with Hill's method for the Floquet analysis of time-periodic solutions of large-scale systems such as those coming from the discretisation of PDEs. }\\